\documentclass[narrowdisplay,onecolumn]{elsart3p}
\usepackage{graphicx}
\usepackage{amssymb}
\usepackage{lineno}

\begin{document}
\hfill {\tiny HISKP-TH-09/24, FZJ-IKP-TH-2009-21}

\begin{frontmatter}
\title{Predictions for the strangeness $S=-3$ and $-4$ baryon-baryon interactions 
in chiral effective field theory}
\author[1]{J. Haidenbauer\corauthref{cor}},
\corauth[cor]{Corresponding author.}
\ead{j.haidenbauer@fz-juelich.de}
\author[1,2]{U.-G. Mei\ss ner}
\address[1]{Institut f{\"u}r Kernphysik, J\"ulich Center for
  Hadron Physics and Institute for Advanced Simulation\\
Forschungszentrum J{\"u}lich, D-52425 J{\"u}lich, Germany}
\address[2]{Helmholtz-Institut f{\"u}r Strahlen- und Kernphysik (Theorie) and
  Bethe Center for Theoretical Physics\\
  Universit{\"a}t Bonn, D-53115 Bonn, Germany}

\begin{abstract}
The leading order strangeness $S=-3$ and $-4$ baryon-baryon interactions are 
analyzed within chiral effective field theory. The chiral potentials consist of contact 
terms without derivatives and of one-pseudoscalar-meson exchanges. 
Assuming ${\rm SU(3)}$ flavor symmetry those contact terms and the couplings of the 
pseudoscalar mesons to the baryons are related to the corresponding quantities of 
the $S=-1$ hyperon-nucleon channels. Specifically, the values of the pertinent
five low-energy constants related to the contact terms are already fixed from our
preceeding study of the $\Lambda N$ and $\Sigma N$ systems and thus
genuine predictions for the $\Xi\Lambda$, $\Xi\Sigma$, and $\Xi\Xi$ interactions 
can be made. Strong attraction is found in some of the $S=-3$ and $-4$ channels, 
suggesting the possible existence of bound states. 

\end{abstract}
\end{frontmatter}


\section{Introduction}
\label{chap:1}

The study of baryon-baryon systems involving strangeness has the potential of 
considerably deepening  our understanding of strong interaction physics. 
Unfortunately, the experimental information on such systems is rather limited. 
Though there is at least a fair amount of {\it concrete} data on the $\Lambda N$ 
and $\Sigma N$ systems, 
the empirical information on the strangeness $S=-2$ sector is only of qualitative
nature. It consists of limits for the $\Xi^-p \to \Xi^-p$ and
$\Xi^-p\to \Lambda\Lambda$ reaction cross sections \cite{Ahn:2005jz} 
and in the binding energy of ${}_{\Lambda\Lambda}^{\;\;\;6}{\rm He}$ \cite{Takahashi:2001nm}, 
where the latter implies that the $\Lambda\Lambda$ interaction can be only moderately attractive.
Virtually nothing is known about the baryon-baryon interaction in the $S=-3$ and $-4$ 
systems. 
Indeed, even theoretical investigations on those systems are rather scarce 
\cite{Stoks:1999bz,Fujiwara:2006yh}.
In 1999, the Nijmegen group presented an extension of their meson-exchange hyperon-nucleon
($YN$; $Y=\Lambda, \ \Sigma$) potential \cite{Rijken1999} 
to interaction channels with $S=-2$, $-3$, and  
$-4$ \cite{Stoks:1999bz} invoking ${\rm SU(3)}$ flavor symmetry arguments. In the actual
model calculation, ${\rm SU(3)_f}$ symmetry is broken by using the physical values
of the involved baryon and meson masses and, in addition, some breaking of the
baryon-baryon-meson coupling constants is allowed.
The interactions predicted in this way
for the $S=-3$ and $-4$ sectors turned out to be fairly strong and attractive and
even suggests the existence of bound states in the $\Xi\Sigma$ and $\Xi\Xi$ channels
\cite{Stoks:1999bz}.
Such strongly attractive $S=-3$ and $-4$ baryon-baryon interactions have further
interesting implications. For example, Filikin and Gal argued that then fairly
light hypernuclei with $S=-3$ should exist \cite{Gal1,Gal2}. In particular, 
${}_{\Xi\Lambda}^{\;\;\;6}{\rm He}$ could then already mark the onset of nuclear
stability for $\Xi$ hyperons. 
A strong $\Xi\Xi$ interaction could induce a first order phase transition from neutron matter
to hyperon-rich matter \cite{Schaffner2008} so that, besides ordinary white dwarfs
and neutron stars, a new class of compact stars, namely hyperon stars, could
exist. 
 
The baryon-baryon interaction of Fujiwara and collaborators \cite{Fujiwara:2006yh}
is derived in the constituent quark model. It contains the color Fermi-Breit 
interaction and effective-meson exchanges (pseudoscalar, scalar, vector mesons)
between quarks, among others. Also here ${\rm SU(3)_f}$ symmetry plays a
key role in extending the model from the $NN$ and $YN$ interaction (where free
parameters are fixed) to the $S=-3$ and $-4$ channels. And also here an
explicit flavor-symmetry breaking is introduced, namely in the Fermi-Breit 
interaction. 
But in this approach it was found that the baryon-baryon interaction becomes 
step by step less attractive when going from strangeness $S=0$ to $S=-4$. 
In particular, no di-baryon bound states are supported, except for the deuteron. 
 
In this context we present here a first study of the baryon-baryon interaction
in the $S=-3$ and $-4$ sectors in the framework of {\em chiral effective field 
theory} (EFT). Outlined by Weinberg in the early 1990s 
\cite{Weinberg:1990rz,Weinberg:1991um}
the concepts of chiral EFT have been applied very successfully in the last decade 
to the $NN$ (and $NNN$) interaction and to the physics of light 
nuclei, resulting in a high-precision description of the experimental data, see e.g. 
Refs.~\cite{Bedaque:2002mn,Epelbaum:2005pn,Epelbaum:2008ga} and references therein. 
Recently, this framework was utilized by us also for investigating the $YN$ 
interactions as well as the baryon-baryon interaction in
the $S=-2$ sector \cite{Polinder:2006zh,Polinder}. In particular, 
we showed that the leading order (LO) chiral EFT successfully describes the available 
$YN$ scattering data \cite{Polinder:2006zh}. Also the binding energies of the light 
hypernuclei are predicted well within chiral EFT \cite{Nogga:2006ir,Haidenbauer:2007ra}. 

The extension of our investigations of the baryon-baryon interaction within
chiral EFT to the strangeness $S=-3$ and $-4$ sectors is straightforward.
The LO potential consists of four-baryon contact terms without derivatives and of 
one-pseudoscalar-meson exchanges \cite{Polinder:2006zh,Haidenbauer:2007ra}. 
Under the assumption of ${\rm SU(3)_f}$ symmetry 
the interaction for the $S=-3$ and $-4$ baryon-baryon sector depends on the
same (five) contact terms that enter also in the $YN$ interaction. 
Thus, we can take over the values which were fixed in our study of the 
$YN$ sector by a fit to the pertinent ($\Lambda N$, $\Sigma N$) data \cite{Polinder:2006zh}. 
Then the interaction in the $S=-3$ and $-4$ channels is a genuine prediction 
that follows from the results of Ref.~\cite{Polinder:2006zh} and 
${\rm SU(3)_f}$ symmetry.

The paper is structured as follows: In Sect. II we provide a short overview of the
chiral EFT approach with emphasis on its extension to the strangeness
$S=-3$ and $-4$ sectors. In Sect. III we show results for the coupled 
$\Xi\Lambda - \Xi\Sigma$ system and for the $\Xi\Xi$ channel. Specifically, 
we present integrated cross sections and effective range parameters for the
$S$-waves and we compare our results to those of the two potential models
mentioned above. The paper ends with a short summary.  


\section{The effective strangeness $S=-3$ and $-4$ baryon-baryon potential}
\label{chap:2}

We construct the chiral effective potentials for the strangeness $S=-3$ and $-4$ 
sector at LO using 
the Weinberg power counting similar to the $YN$ case considered in \cite{Polinder:2006zh}. 
The LO potential consists of four-baryon contact terms without derivatives and of 
one-pseudoscalar-meson exchanges. 
Details on the derivation of the LO contact terms for the octet baryon-baryon interaction
can be found in Ref. \cite{Polinder:2006zh,Haidenbauer:2007ra}. Here we only
list the final result for the $\Xi Y$ and $\Xi\Xi$ partial wave potentials in
the singlet and triplet S-waves, cf. Table~\ref{tab:2.1}. 
Also the $S=0$ and $-1$ potentials are included in Table~\ref{tab:2.1} for completeness. 
For convenience in the present paper we express the 
baryon-baryon potentials in terms of the ${\rm SU(3)_f}$ irreducible 
representations, see e.g. \cite{deSwart:1963gc} and also \cite{Iwao,Dover:1991sh},
so that the contact interaction is given by 
\begin{equation}
V^{B_1B_2\to B_1'B'_2}=
\frac{1}{4}(1-\mbox{\boldmath $\sigma$}_1\cdot \mbox{\boldmath $\sigma$}_2) \, C_{1S0}
+ \frac{1}{4}(3+\mbox{\boldmath $\sigma$}_1 \cdot\mbox{\boldmath $\sigma$}_2) \, C_{3S1} \ . 
\label{contact}
\end{equation}
From the relations in Table~\ref{tab:2.1} we can see that for the ${}^1S_0$ channel 
there is a
one-to-one correspondence between the $S=0$ and $S=-4$ sectors, i.e. the $NN$ 
and the $\Xi\Xi$ interactions, and also between the $S=-1$ and $S=-3$ sectors,
i.e. the $YN$ and $\Xi Y$ interactions. Thus, if ${\rm SU(3)_f}$ symmetry
is fully realized then not only the potentials but even the reaction amplitudes
would be the same in the corresponding sectors. Of course, the symmetry is no
longer fulfilled once physical values for the baryon masses are used in the
evaluation of the scattering observables, even for a 
${\rm SU(3)_f}$ symmetric interaction potential. 
Note that 
there is no such correspondence for the ${}^3S_1$ channel because here the
role of the $10$ and $10^*$ representations is interchanged when going from
$S=0$ to $S=-4$ and from $S=-1$ to $S=-3$, respectively.
\begin{table}[t]
\caption{Various LO baryon-baryon contact potentials for the ${}^1S_0$ and ${}^3S_1$ partial 
waves in the isospin basis. These potentials are flavor symmetric. $C^{27}$ etc. refers to 
the corresponding ${\rm SU(3)_f}$ irreducible representation \cite{deSwart:1963gc,Dover:1991sh}. 
}
\label{tab:2.1}
\vspace{0.2cm}
\centering
\begin{tabular}{|l|c|c|l|c|l|}
\hline
&Channel &Isospin &$C_{1S0}$ &Isospin &$C_{3S1}$\\
\hline
$S=0$&$NN\rightarrow NN$ &$1$ & $C^{27}$ &$0$ &$C^{10^*}$\\
\hline
$S=-1$&$\Lambda N \rightarrow \Lambda N$ &$\frac{1}{2}$ &$\frac{1}{10}\left(9C^{27}+C^{8_s}\right)$
&$\frac{1}{2}$ &$\frac{1}{2}\left(C^{8_a}+C^{10^*}\right)$\\
&$\Lambda N \rightarrow \Sigma N$ &$\frac{1}{2}$ &$\frac{3}{10}\left(-C^{27}+C^{8_s}\right)$
&$\frac{1}{2}$ &$\frac{1}{2}\left(-C^{8_a}+C^{10^*}\right)$\\
&$\Sigma N \rightarrow \Sigma N$ &$\frac{1}{2}$ &$\frac{1}{10}\left(C^{27}+9C^{8_s}\right)$
&$\frac{1}{2}$ &$\frac{1}{2}\left(C^{8_a}+C^{10^*}\right)$\\
&$\Sigma N \rightarrow \Sigma N$ &$\frac{3}{2}$ &$C^{27}$
&$\frac{3}{2}$ &$C^{10}$\\
\hline
$S=-3$&$\Xi\Lambda \rightarrow \Xi\Lambda $ &$\frac{1}{2}$ &$\frac{1}{10}\left(9C^{27}+C^{8_s}\right)$
&$\frac{1}{2}$ &$\frac{1}{2}\left(C^{8_a}+C^{10}\right)$\\
&$\Xi \Lambda \rightarrow \Xi \Sigma $ &$\frac{1}{2}$ &$\frac{3}{10}\left(-C^{27}+C^{8_s}\right)$
&$\frac{1}{2}$ &$\frac{1}{2}\left(-C^{8_a}+C^{10}\right)$\\
&$\Xi \Sigma \rightarrow \Xi \Sigma $ &$\frac{1}{2}$ &$\frac{1}{10}\left(C^{27}+9C^{8_s}\right)$
&$\frac{1}{2}$ &$\frac{1}{2}\left(C^{8_a}+C^{10}\right)$\\
&$\Xi \Sigma \rightarrow \Xi \Sigma $ &$\frac{3}{2}$ &$C^{27}$
&$\frac{3}{2}$ &$C^{10^*}$\\
\hline
$S=-4$&$\Xi\Xi\rightarrow \Xi\Xi$ &$1$ & $C^{27}$ &$0$ &$C^{10}$\\
\hline
\end{tabular}
\end{table}

The actual values of the $S$-wave contact terms corresponding to the irreducible
representations of ${\rm SU(3)_f}$ are summarized in Table~\ref{tab:6.1} for the
various cut--offs (as defined more precisely below). These values were fixed
by a fit to the $YN$ date in
our earlier study \cite{Polinder:2006zh}. They are obtained from the low-energy
constants listed in Table~2 of that paper via the relations Eqs.~(2.8) - (2.13) 
given therein as well.

\begin{table}[t]
\caption{The $S$-wave contact terms corresponding to the irreducible 
representations of ${\rm SU(3)_f}$ for various cut--offs.
The values of
the LECs are in $10^4$ ${\rm GeV}^{-2}$; the values of $\Lambda$ are in MeV.
}
\label{tab:6.1}
\vspace{0.2cm}
\centering
\begin{tabular}{|l|rrrr|}
\hline
$\ \Lambda$& $550$& $600$& $650$& $700 \ $  \\
\hline
$C^{27}$   &$-.0766$ &$-.0763$ &$-.0757$ &$-.0744$\\
$C^{10^*}$ &$-.0239$ &$-.0182$ &$-.0097$ &$ .0013$\\
$C^{10}$   &$.2336$  &$.2391$  &$.2392$  &$.2501$\\
$C^{8_s}$  &$.2241$ &$.2839$ &$.3595$ &$.3650$\\
$C^{8_a}$  &$-.0206$  &$-.0144$  &$-.0097$  &$-.0057$\\
\hline
\end{tabular}
\end{table}

The lowest order ${\rm SU(3)_f}$ invariant pseudoscalar-meson--baryon
interaction Lagrangian with the appropriate symmetries was discussed in 
\cite{Polinder:2006zh}. In the isospin basis it reads
\begin{eqnarray}
{\mathcal L}&=&-f_{NN\pi}\bar{N}\gamma^\mu\gamma_5\mbox{\boldmath $\tau$}N\cdot\partial_\mu\mbox{\boldmath $\pi$} +if_{\Sigma\Sigma\pi}\bar{\mbox{\boldmath $ \Sigma$}}\gamma^\mu\gamma_5\times{\mbox{\boldmath $ \Sigma$}}\cdot\partial_\mu\mbox{\boldmath $\pi$} \nonumber \\
&&-f_{\Lambda\Sigma\pi}\left[\bar{\Lambda}\gamma^\mu\gamma_5{\mbox{\boldmath $ \Sigma$}}+\bar{\mbox{\boldmath $\Sigma$}}\gamma^\mu\gamma_5\Lambda\right]\cdot\partial_\mu\mbox{\boldmath $\pi$}-f_{\Xi\Xi\pi}\bar{\Xi}\gamma^\mu\gamma_5\mbox{\boldmath $\tau$}\Xi\cdot\partial_\mu\mbox{\boldmath $\pi$} \nonumber \\
&&-f_{\Lambda NK}\left[\bar{N}\gamma^\mu\gamma_5\Lambda\partial_\mu K+\bar{\Lambda}\gamma^\mu\gamma_5N\partial_\mu K^\dagger\right]
-f_{\Xi\Lambda K}\left[\bar{\Xi}\gamma^\mu\gamma_5\Lambda\partial_\mu K_c+\bar{\Lambda}\gamma^\mu\gamma_5\Xi\partial_\mu K_c^\dagger\right]
\nonumber \\&&
-f_{\Sigma NK}\left[\bar{\mbox{\boldmath $ \Sigma$}}\cdot\gamma^\mu\gamma_5\partial_\mu K^\dagger\mbox{\boldmath $\tau$}N+\bar{N}\gamma^\mu\gamma_5\mbox{\boldmath $\tau$}\partial_\mu K\cdot{\mbox{\boldmath $ \Sigma$}}\right]
-f_{\Xi \Sigma K}\left[\bar{\mbox{\boldmath $ \Sigma$}}\cdot\gamma^\mu\gamma_5\partial_\mu K_c^\dagger\mbox{\boldmath $\tau$}\Xi+\bar{\Xi}\gamma^\mu\gamma_5\mbox{\boldmath $\tau$}\partial_\mu K_c\cdot{\mbox{\boldmath $ \Sigma$}}\right]
\nonumber \\&&
-f_{NN\eta_8}\bar{N}\gamma^\mu\gamma_5N\partial_\mu\eta
-f_{\Lambda\Lambda\eta_8}\bar{\Lambda}\gamma^\mu\gamma_5\Lambda\partial_\mu\eta-f_{\Sigma\Sigma\eta_8}\bar{\mbox{\boldmath $ \Sigma$}}\cdot\gamma^\mu\gamma_5{\mbox{\boldmath $ \Sigma$}}\partial_\mu\eta
-f_{\Xi\Xi\eta_8}\bar{\Xi}\gamma^\mu\gamma_5\Xi\partial_\mu\eta \ .
\label{eq:2.3}
\end{eqnarray}
The interaction Lagrangian in Eq.~(\ref{eq:2.3}) is invariant under ${\rm SU(3)_f}$ 
transformations if the various coupling constants fulfill specific relations which 
can be expressed in terms of the coupling constant $f$ and the $F/(F+D)$-ratio $\alpha$ 
as \cite{deSwart:1963gc}, 
\begin{equation}
\begin{array}{rlrlrl}
f_{NN\pi}  = & f, & f_{NN\eta_8}  = & \frac{1}{\sqrt{3}}(4\alpha -1)f, & f_{\Lambda NK} = & -\frac{1}{\sqrt{3}}(1+2\alpha)f, \\
f_{\Xi\Xi\pi}  = & -(1-2\alpha)f, &  f_{\Xi\Xi\eta_8}  = & -\frac{1}{\sqrt{3}}(1+2\alpha )f, & f_{\Xi\Lambda K} = & \frac{1}{\sqrt{3}}(4\alpha-1)f, \\
f_{\Lambda\Sigma\pi}  = & \frac{2}{\sqrt{3}}(1-\alpha)f, & f_{\Sigma\Sigma\eta_8}  = & \frac{2}{\sqrt{3}}(1-\alpha )f, & f_{\Sigma NK} = & (1-2\alpha)f, \\
f_{\Sigma\Sigma\pi}  = & 2\alpha f, &  f_{\Lambda\Lambda\eta_8}  = & -\frac{2}{\sqrt{3}}(1-\alpha )f, & f_{\Xi\Sigma K} = & -f.
\end{array}
\label{eq:2.5}
\end{equation}
Here $f\equiv g_A/2F_\pi$, $g_A$ is the axial-vector strength, $g_A= 1.26$, which is 
measured in neutron $\beta$--decay and $F_\pi$ is the weak pion decay constant, 
$F_\pi =  92.4$ MeV. For the $F/(F+D)$-ratio we adopt here the ${\rm SU(6)}$
value ($\alpha=0.4$) 
which was already used in our study of the $YN$ system \cite{Polinder:2006zh}. The 
spin-space part of the one-pseudoscalar-meson-exchange potential resulting from the 
interaction Lagrangian Eq.~(\ref{eq:2.3}) is in leading order similar to the static 
one-pion-exchange potential in Ref.~\cite{Epelbaum:1998ka},
\begin{eqnarray}
V^{B_1B_2\to B_1'B'_2}&=&-f_{B_1B'_1P}f_{B_2B'_2P}
\frac{\left(\mbox{\boldmath $\sigma$}_1\cdot{\bf k}\right)\left(\mbox{\boldmath $
\sigma$}_2\cdot{\bf k}\right)}{{\bf k}^2+m^2_P}\ , 
\label{eq:2.6}
\end{eqnarray}
where $f_{B_1B'_1P}$, $f_{B_2B'_2P}$ are the appropriate coupling constants as 
given in Eq.~(\ref{eq:2.5}) and $m_P$ is the actual mass of the exchanged pseudoscalar 
meson. With regard to the $\eta$ meson we identified its coupling with the octet value, 
i.e. the one for $\eta_8$. The transferred momentum ${\bf k}$ is defined 
in terms of the final and initial center-of-mass (c.m.) momenta of the 
baryons, ${\bf p}_f$ and ${\bf p}_i$, as ${\bf k}={\bf p}_f-{\bf p}_i$.
To find the complete LO one-pseudoscalar-meson-exchange 
potential one needs to multiply the potential in Eq.~(\ref{eq:2.6}) with the isospin 
factors ${\mathcal I}$ given in Table~\ref{tab:3.1}.
\begin{table}
\caption{Isospin factors ${\mathcal I}$ for the various one--pseudoscalar-meson exchanges.
}
\label{tab:3.1}
\centering
\begin{tabular}{|l|c|c|c|c|}
\hline
Channel &Isospin &$\ \pi \ $ &$\ K \ $ &$\ \eta \ $\\
\hline
$\Xi\Xi\to \Xi\Xi$ &$0$ &$-3$ &$0$ &$1$ \\
                   &$1$ &$1$  &$0$ &$1$ \\
\hline
$\Xi\Lambda \to \Xi\Lambda $ &$\frac{1}{2}$ &$0$ &$1$ &$1$ \\
\hline
$\Xi\Lambda \to \Xi\Sigma $ &$\frac{1}{2}$ &$-\sqrt{3}$ &$-\sqrt{3}$ &$0$ \\
\hline
$\Xi\Sigma \to \Xi\Sigma $ &$\frac{1}{2}$ &$-2$ &$-1$ &$1$ \\
                               &$\frac{3}{2}$ &$1$ &$2$ &$1$ \\
\hline
\end{tabular}
\end{table}

The ${\rm SU(3)_f}$ symmetry of the one-pseudoscalar-meson exchanges is 
broken by the masses of the pseudoscalar mesons. This is taken into account explicitly 
in Eq.~(\ref{eq:2.6}) by taking the appropriate values for $m_P$. In case one would 
consider identical pseudoscalar-meson masses, the corresponding potential obeys the 
${\rm SU(3)_f}$ relations as shown in the 4th and 6th column of Table~\ref{tab:2.1},
respectively. 

Finally, for completeness we briefly comment on the used scattering equation. The 
calculations are done in momentum space. We solve the coupled--channel
(non--relativistic) Lippmann-Schwinger (LS) equation,
\begin{eqnarray}
T_{\rho''\rho'}^{\nu''\nu',J}(p'',p';\sqrt{s})&=&V_{\rho''\rho'}^{\nu''\nu',J}(p'',p')+
\sum_{\rho,\nu}\int_0^\infty \frac{dpp^2}{(2\pi)^3} \, V_{\rho''\rho}^{\nu''\nu,J}(p'',p) 
\frac{2\mu_{\nu}}{q_{\nu}^2-p^2+i\epsilon}T_{\rho\rho'}^{\nu\nu',J}(p,p';\sqrt{s})\ .
\end{eqnarray}
The label $\nu$ indicates the particle channels and the label $\rho$ the partial 
wave. $\mu_\nu$ is the pertinent reduced mass. The on-shell momentum in the 
intermediate state, $q_{\nu}$, is defined by 
$\sqrt{s}=\sqrt{M^2_{B_{1,\nu}}+q_{\nu}^2}+\sqrt{M^2_{B_{2,\nu}}+q_{\nu}^2}$. 
Relativistic kinematics is used for relating the laboratory energy $T_{{\rm lab}}$ 
of the baryons to the center-of-mass momentum. Suppressing the particle channel label, 
the partial wave projected potentials $V_{\rho''\rho'}^{J}(p'',p')$ are given in 
\cite{Polinder:2006zh}. The LS equation for the coupled channels 
$\Xi \Lambda$ and $\Xi \Sigma$ is solved in the particle basis, in 
order to incorporate the correct physical thresholds. The potential in the LS 
equation is cut off with the regulator function $f^\Lambda(p',p)$,
\begin{equation}
f^\Lambda(p',p)=e^{-\left(p'^4+p^4\right)/\Lambda^4}\ ,
\end{equation}
in order to remove high-energy components of the baryon and pseudoscalar meson fields. 
We consider again cut-off values in the range of 550$\ldots$700 MeV as in 
Ref.~\cite{Polinder:2006zh}. This range is also similar to the one considered in 
the $NN$ case, see, e.g. Refs.~\cite{Epelbaum:2002ji,Epelbaum:2003xx}. 
The cross sections are calculated using the (LSJ basis) partial wave amplitudes, 
for details we refer to \cite{Nag75,Holzenkamp:1989tq}. 


\section{Results and discussion}
\label{chap:4}

The LO chiral EFT interaction for the $S=-3$ and $-4$ baryon-baryon sector depends only 
on those five contact terms that enter also in the $YN$ interaction. Thus, we 
can take over the values which were fixed in our study of the $YN$ sector \cite{Polinder:2006zh}. 
Then the interaction in the $S=-3$ and $-4$ channels is a genuine prediction that follows
from the results of Ref.~\cite{Polinder:2006zh} and ${\rm SU(3)_f}$ symmetry.

\begin{figure}[t]
\begin{center}
\includegraphics[height=73mm]{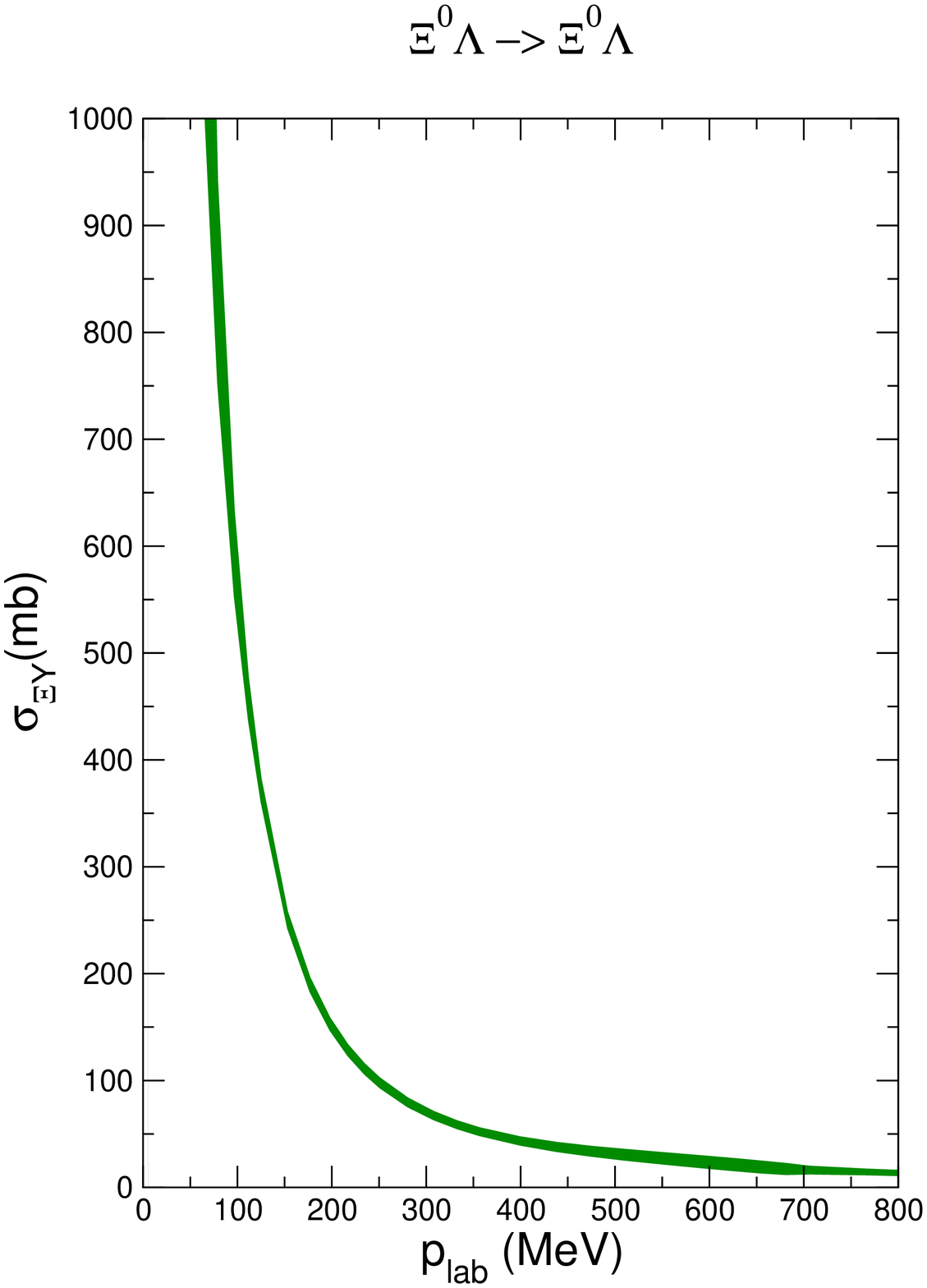}
\includegraphics[height=73mm]{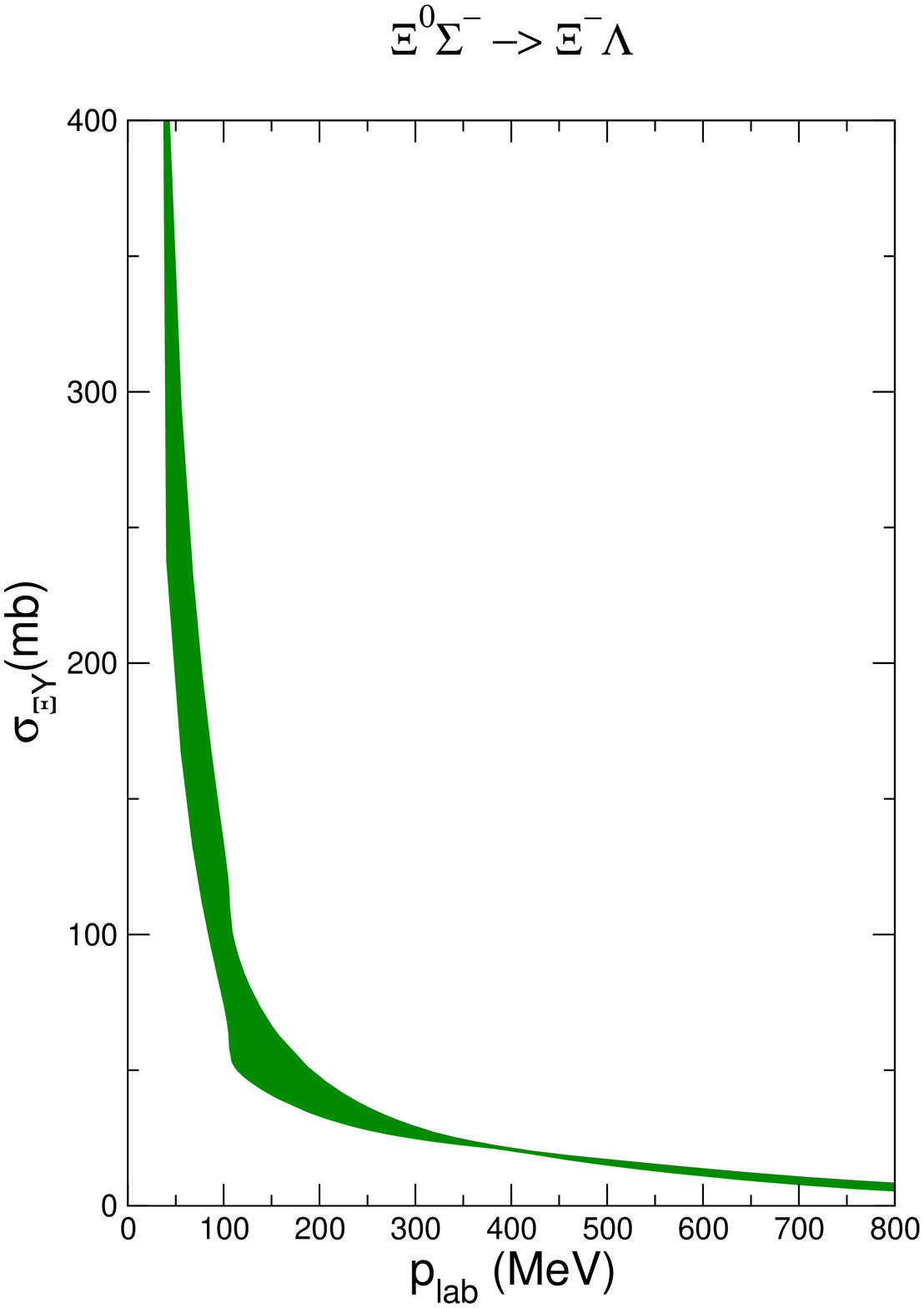}
\includegraphics[height=73mm]{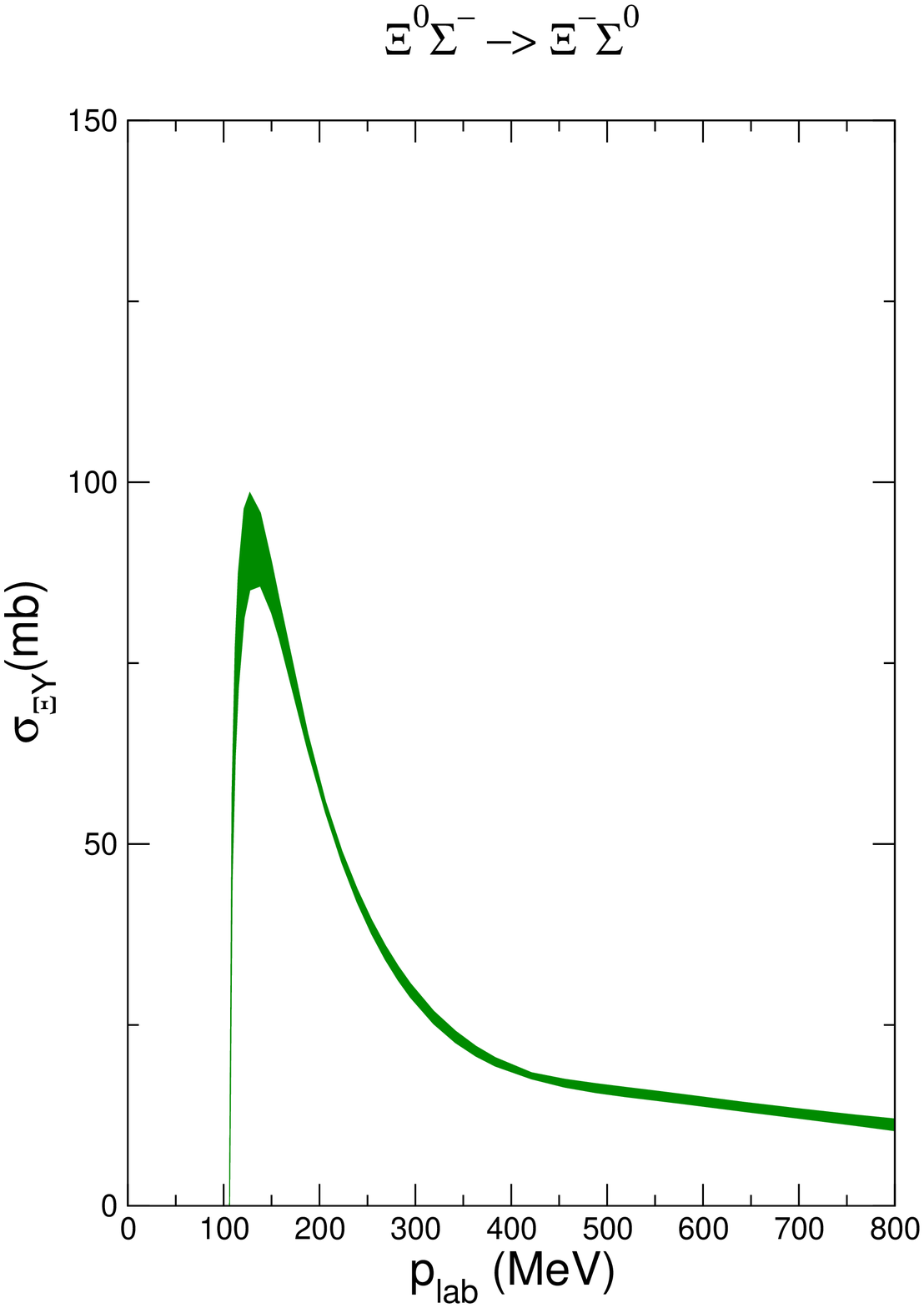}

\includegraphics[height=73mm]{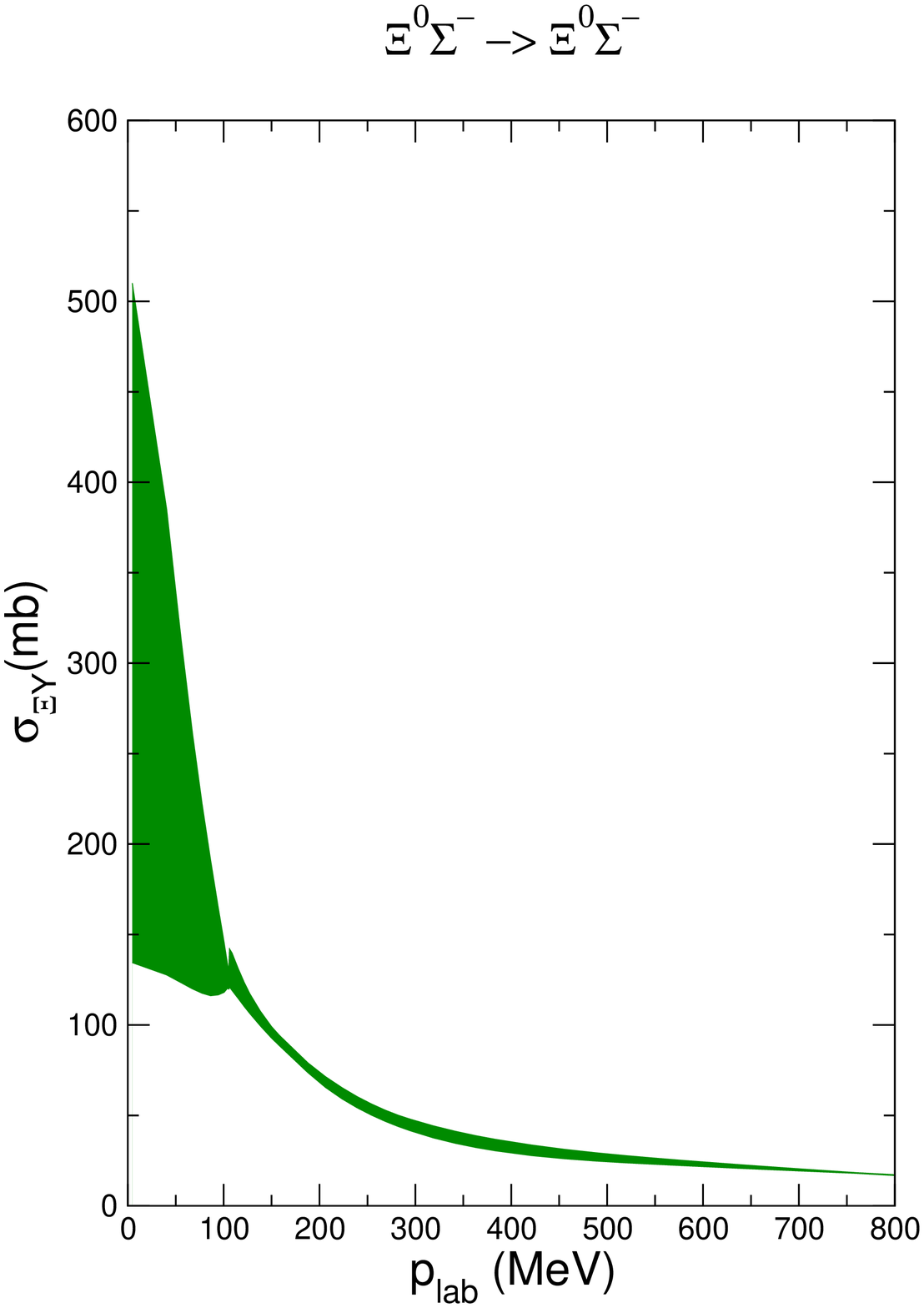}
\includegraphics[height=73mm]{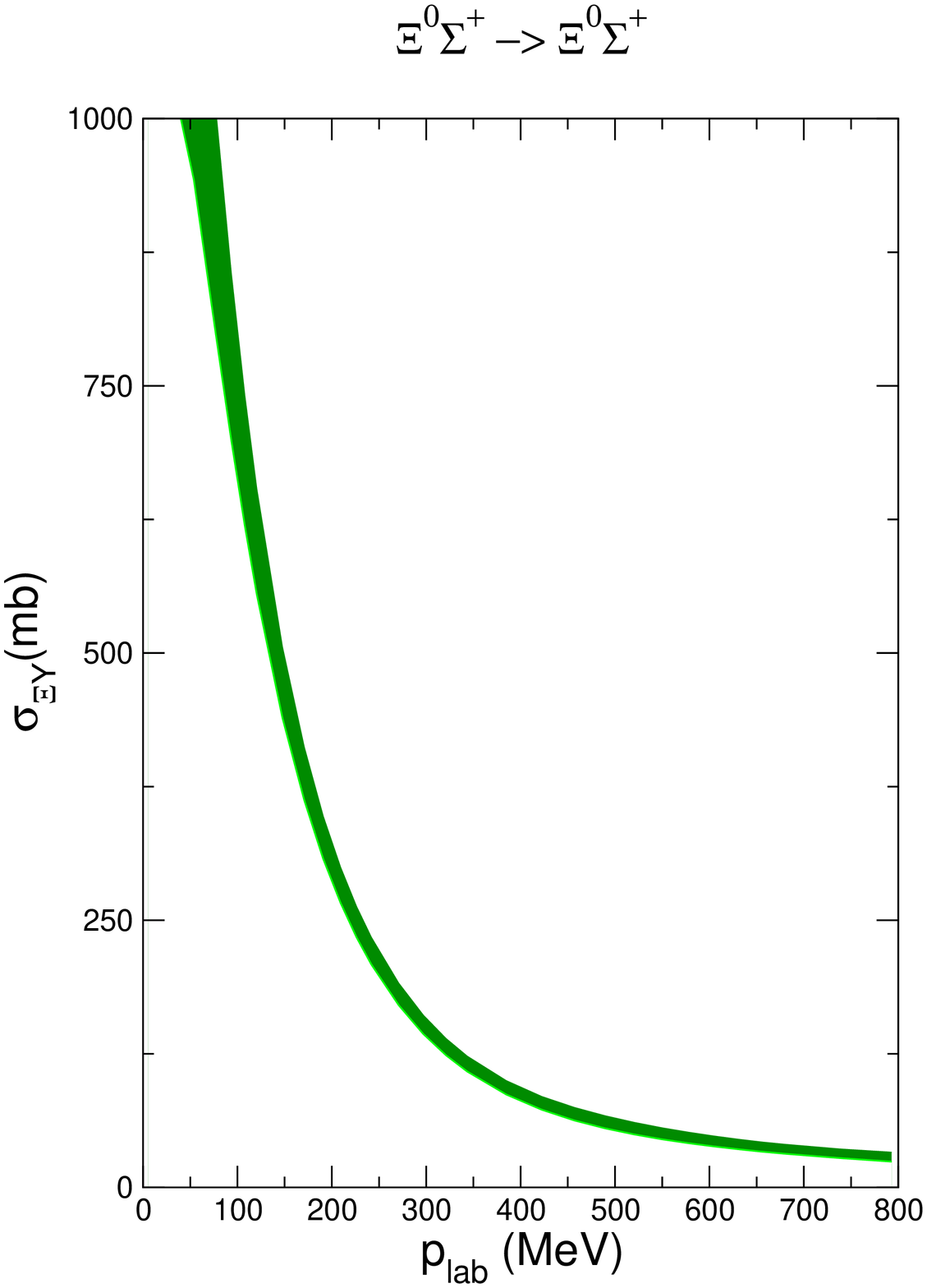}
\caption{Total cross sections for various reactions in the strangeness $S=-3$ sector 
as a function of  $p_{lab}$.
The shaded band shows the chiral EFT results for variations of the cut-off
in the range $\Lambda =$ 550$\ldots$700~MeV.
}
\label{fig:4.2}
\end{center}
\end{figure}

Results for the $\Xi^0\Lambda \to \Xi^0\Lambda$, $\Xi^0\Sigma^-\to \Xi^-\Lambda$,
$\Xi^0\Sigma^-\to \Xi^-\Sigma^0$, $\Xi^0\Sigma^-\to \Xi^0\Sigma^-$, and 
$\Xi^0\Sigma^+\to \Xi^0\Sigma^+$ scattering cross sections are shown 
in Fig.~\ref{fig:4.2}. Partial waves with total angular momentum 
up-to-and-including $J = 2$ are taken into account. 
The shaded bands show the cut-off dependence. 
{}From that figure one observes that the $\Xi^0\Lambda \to \Xi^0\Lambda$
and $\Xi^0\Sigma^+\to \Xi^0\Sigma^+$ cross sections are rather large near threshold. 
Though the cross section for $\Xi^0\Sigma^-\to \Xi^-\Lambda$ rises too, 
in this case it is only due to the phase space factor 
$p_{\Xi^-\Lambda}/p_{\Xi^0\Sigma^-}$. 
There is a clear cusp effect visible in 
the $\Xi^0\Sigma^-$ cross section at $p_{lab} \approx 106$ MeV/c, i.e. at
the opening of the $\Xi^-\Sigma^0$ channel. On the other hand, we do not observe
any sizeable cusp effects in the $\Xi^0\Lambda$ cross section around $p_{lab} = 690$ 
MeV/c, i.e. at the opening of the $\Xi\Sigma$ channels. The latter is in line with 
the results reported by the Nijmegen group for their interactions,
where a cusp effect in that channel is absent too. 
In this context we would like to remind the reader that the cusp seen in the 
corresponding strangeness $S=-2$ case, namely in the 
$\Lambda N$ cross section at the $\Sigma N$ threshold,  
is rather pronounced in our chiral EFT interaction \cite{Polinder:2006zh} 
but also in conventional meson-exchange potential models \cite{Rijken1999,Haidenbauer}.

\begin{figure}[t]
\begin{center}
\includegraphics[height=73mm]{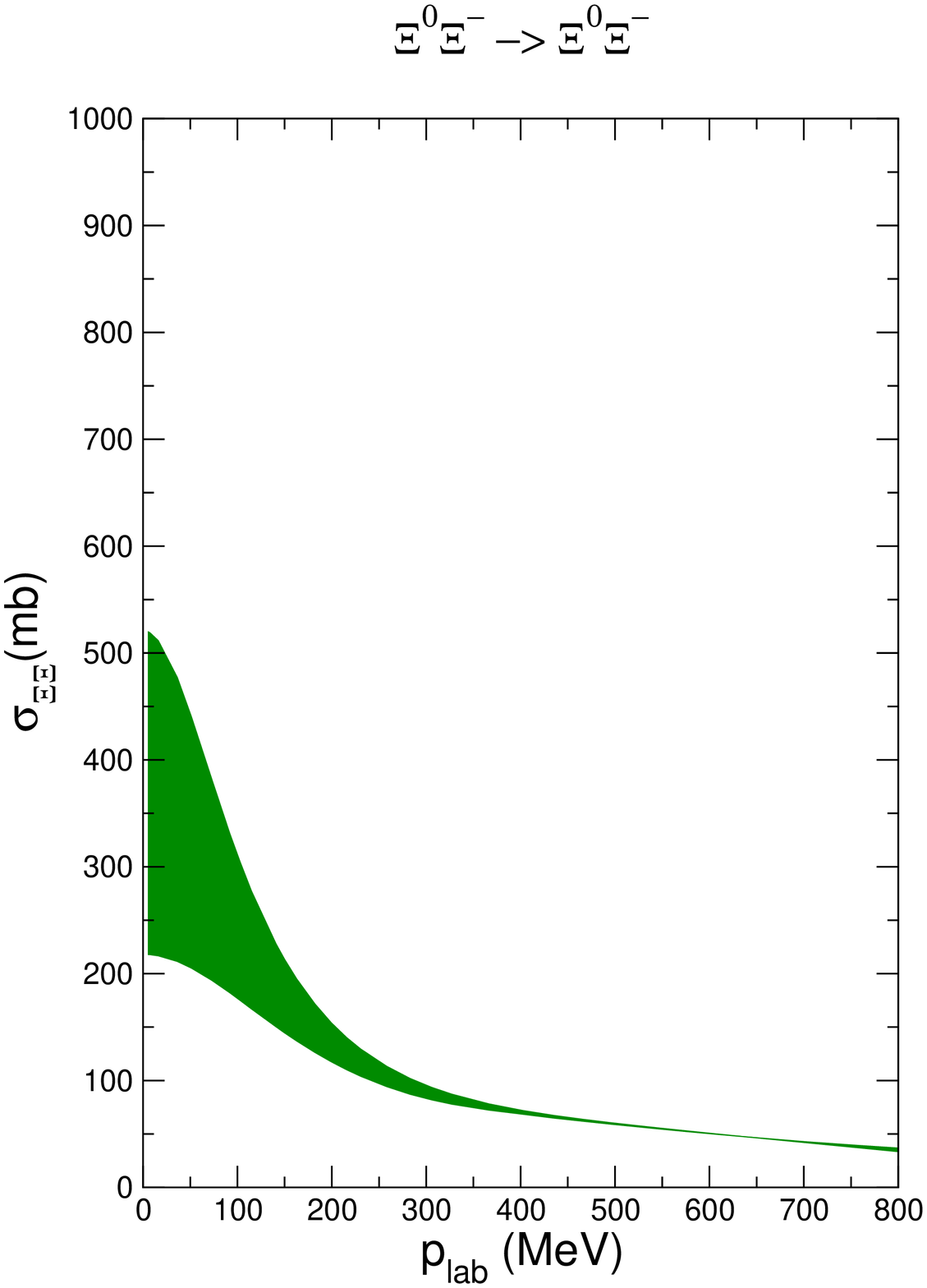}
\includegraphics[height=73mm]{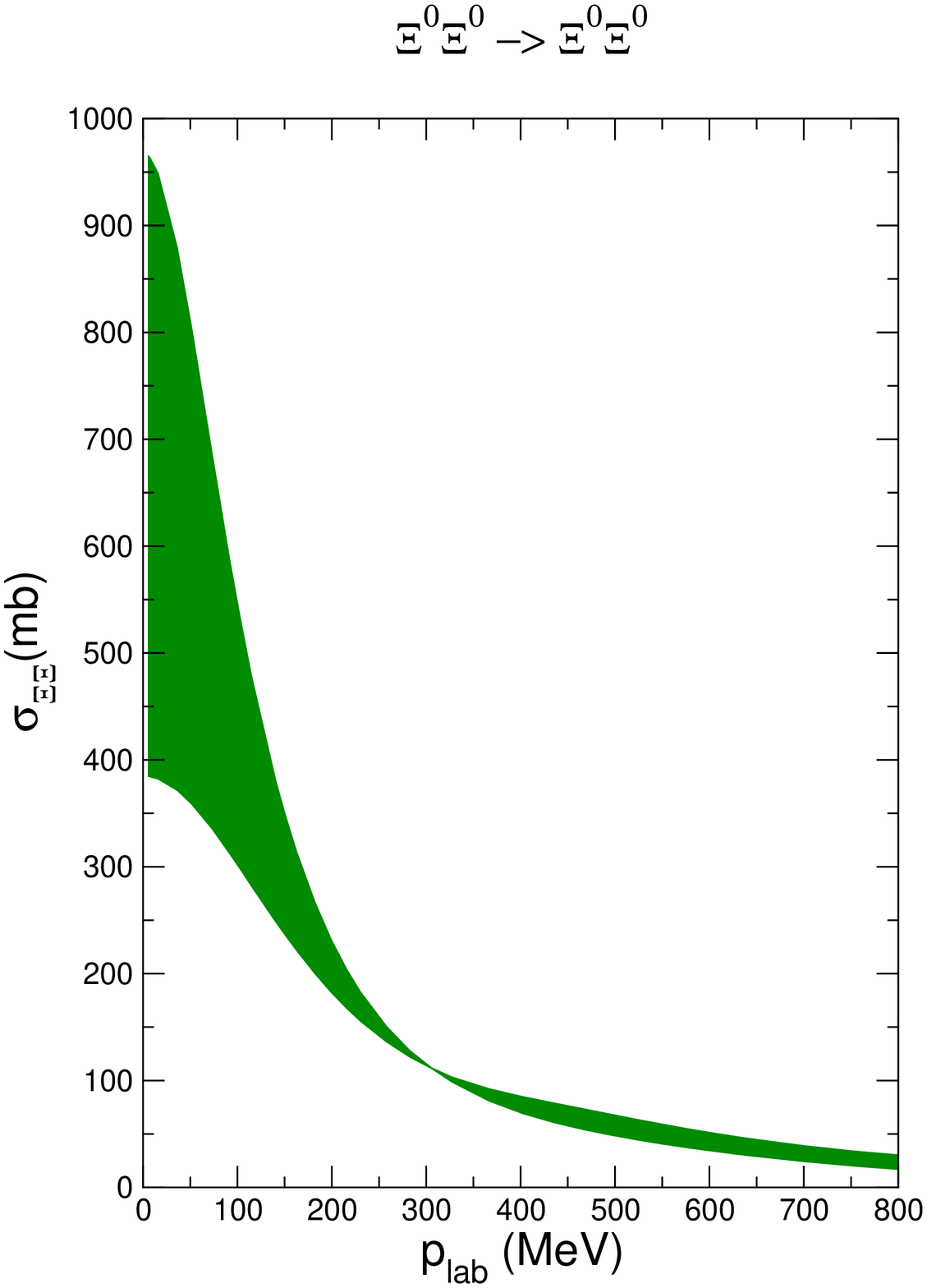}
\caption{Total cross sections for the reactions 
$\Xi^0\Xi^-\to \Xi^0\Xi^-$ and $\Xi^0\Xi^0\to \Xi^0\Xi^0$
as a function of $p_{lab}$.
The shaded band shows the chiral EFT results for variations of the cut-off
in the range $\Lambda =$ 550$\ldots$700 MeV.
}
\label{fig:4.3}
\end{center}
\end{figure}

Cross section results for the $\Xi^0\Xi^0$ and $\Xi^0\Xi^-$ channels are shown in 
Fig.~\ref{fig:4.3}, again as a function of $p_{lab}$ and with 
shaded bands that indicate the cut-off dependence. 

Results for scattering cross sections at $p_{lab} =$~150~MeV/c
for the various $\Xi Y$ and $\Xi\Xi$ channels are listed in 
Table~\ref{tab:r1}. We also include predictions by other models
\cite{Stoks:1999bz,Fujiwara:2006yh} for channels where pertinent results are
available in the literature. Those cross sections were evaluated from the 
effective range parameters given in the corresponding publications. 
 
The table makes clear again that our predictions for the $\Xi^0\Lambda$ cross section
are indeed sizeable. The results of the Nijmegen meson-exchange
potential and of the quark model of Fujiwara et al. are significantly smaller in that
channel. But we would like to remark that at least one of the interactions
from the Nijmegen group (NSC97b) yields  values that are already fairly close
to ours (with $\sigma_{\Xi^0 \Lambda} \approx 250\,$mb at $p_{lab} =$~150~MeV/c). 
In the $\Xi^0\Sigma^+$ channel (which is a pure isospin $I=3/2$ state) the cross 
sections are also rather large. But here the predictions from the other interactions
considered are of comparable magnitude. 

It is reassuring to see that the variation of our results with the cut-off mass 
is not very pronounced. In fact, in general the differences in the cross sections 
are not more than 20\% at $p_{lab} =$~150~MeV/c and, thus, exhibit uncertainties 
very similar to those that we have observed in our analysis of the 
$\Lambda N$ and $\Sigma N$ cross sections \cite{Polinder:2006zh}.
In this context let us mention that Stoks and Rijken noticed much more
pronounced differences between their six NSC97 models when going to the 
$S=-3$ and $-4$ sectors, as reported in \cite{Stoks:1999bz}. Their cross sections
often differ by a factor 2 or 3, or even more. But one should keep in mind that 
in the Nijmegen potential an ${\rm SU(3)}$ breaking, in addition to the one induced 
by the mass differences of the pseudoscalar mesons, was introduced, cf. 
Ref.~\cite{Rijken1999}. 

\begin{table}[t]
\caption{The $\Xi Y$ and $\Xi\Xi$ integrated cross sections 
(im mb) at $p_{lab} =$~150~MeV/c for various cut-off values $\Lambda$. 
The last columns show results for the Nijmegen potential (NSC97a, NSC97f) \cite{Stoks:1999bz} 
and the model by Fujiwara et al. (fss2) \cite{Fujiwara:2006yh}. 
}
\label{tab:r1}
\vspace{0.2cm}
\centering
\begin{tabular}{|c|cccc|cc|c|}
\hline
& \multicolumn{4}{|c|}{EFT} & \ NSC97a \ & \ NSC97f \ & \ fss2 \ \\
\hline
${\Lambda}$ (MeV)  & \ 550 \ & \ 600 \ & \ 650 \ & \ 700 \ & & & \\
\hline
$\Xi^0\Lambda\phantom{-}\to\Xi^0\Lambda\phantom{-}$ &$254$ &$269$ &$268$ &$262$ & $38.5$ & $66.7$ & $29.0$\\
$\Xi^0 \Sigma^- \to \Xi^- \Lambda$ &$66.4$ &$58.8$ &$49.5$ &$40.9$ & $$ & $$ & $$\\
$\Xi^0 \Sigma^- \to \Xi^- \Sigma^0$ &$82.0$ &$83.0$ &$85.2$ &$89.0$ & $$ & $$ & $$\\
$\Xi^0 \Sigma^- \to \Xi^0 \Sigma^-$ &$92.8$ &$99.0$ &$99.0$ &$96.8$ & $$ & $$ & $$\\
$\Xi^0 \Sigma^+ \to \Xi^0 \Sigma^+$ &$430$ &$469$ &$484$ &$492$ & $725$ & $357$ & $400$\\
\hline
\hline
$\Xi^0\Xi^0 \to \Xi^0\Xi^0$ &$351$ &$302$ &$261$ &$237$ & $420$ & $250$ & $62.3$\\
$\Xi^0\Xi^- \to \Xi^0\Xi^-$ &$231$ &$184$ &$160$ &$145$ & $228$ & $151$ & $437$\\
\hline
\end{tabular}
\end{table}

Results for the $\Xi^0\Lambda$, $\Xi^0\Sigma^+$, and $\Xi\Xi$ 
scattering lengths and effective ranges are listed in Table~\ref{tab:r2}. Also here
values from the $S=-3$ and $-4$ baryon-baryon interaction potentials of 
Refs.~\cite{Stoks:1999bz,Fujiwara:2006yh} are included. 
\begin{table}[t]
\caption{Selected $\Xi Y$ and $\Xi\Xi$ 
singlet and triplet scattering lengths $a$ and effective ranges $r$ 
(in fm) for various cut-off values $\Lambda$. 
The last columns show results for the Nijmegen potential (NSC97a, NSC97f) \cite{Stoks:1999bz} 
and the model by Fujiwara et al. (fss2) \cite{Fujiwara:2006yh}. 
}
\label{tab:r2}
\vspace{0.2cm}
\centering
\begin{tabular}{|c|cccc|cc|c|}
\hline
& \multicolumn{4}{|c|}{EFT} & \ NSC97a \ & \ NSC97f \ & \ fss2 \ \\
\hline
${\Lambda}$ (MeV)  & \ 550 \ & \ 600 \ & \ 650 \ & \ 700 \ & & & \\
\hline
$a^{\Xi\Lambda}_s$  &$-33.5$ &$35.4$ &$12.7$ &$9.07$ & $-0.80$ & $-2.11$ & $-1.08$\\
$r^{\Xi\Lambda}_s$  &$1.00$  &$0.93$  &$0.87$  &$0.84$ & $4.71$ & $3.21$ & $3.55$\\
$a^{\Xi\Lambda}_t$  &$0.33$  &$0.33$ &$0.32$ &$0.31$ & $0.54$ & $0.33$ & $0.26$\\
$r^{\Xi\Lambda}_t$  &$-0.36$  &$-0.30$  &$-0.29$  &$-0.27$ & $-0.47$ & $2.79$ &$2.15$\\
\hline
$a^{\Xi^0\Sigma^+}_s$  &$4.28$ &$3.45$  &$2.97$  &$2.74$ & $4.13$ & $2.32$ & $-4.63$\\
$r^{\Xi^0\Sigma^+}_s$  &$0.96$  &$0.90$  &$0.84$   &$0.81$ & $1.46$ & $1.17$ & $2.39$\\
$a^{\Xi^0\Sigma^+}_t$  &$-2.45$  &$-3.11$ &$-3.57$   &$-3.89$ & $3.21$ & $1.71$ & $-3.48$\\
$r^{\Xi^0\Sigma^+}_t$  &$1.84$ &$1.72$ & $1.70$ & $1.70$ & $1.28$ & $0.96$ & $2.52$\\
\hline
\hline  
$a^{\Xi\Xi}_s$  &$3.92$ &$3.16$ &$2.71$ &$2.47$ & $17.28$ & $2.38$ & $-1.43$\\
$r^{\Xi\Xi}_s$  &$0.92$  &$0.85$  &$0.79$  &$0.75$ & $1.85$ & $1.29$ & $3.20$\\
$a^{\Xi\Xi}_t$  &$0.63$ &$0.59$   &$0.55$ &$0.52$ &  $0.40$ & $0.48$ & $3.20$\\
$r^{\Xi\Xi}_t$  &$1.04$  &$1.05$  &$1.08$ &$1.11$ & $3.45$ & $2.80$ &$0.22$\\
\hline
\end{tabular}
\end{table}
This table reveals the reason for the sizeable $\Xi^0\Lambda$ cross
section predicted by the chiral EFT interactions, 
namely a rather large scattering length in the corresponding 
${}^1S_0$ partial wave. It is obvious that its value is 
strongly sensitive to cut-off variations. It even changes sign (in other
words, it becomes infinite) within the considered cut-off range. This
means that a virtual bound state transforms into a real bound state,
where the strongest binding occurs for the cut-off 
$\Lambda = 700$ MeV and leads to a binding energy of $-0.43\,$MeV.
While this behaviour is interesting per se, one certainly has to stress
that in such a case the predictive power of our LO calculation is 
rather limited. One has to wait for at least an NLO calculation, where we
expect that the cut-off dependence will become much weaker so that more 
reliable conclusions on the possible existence of a virtual or a real 
bound state should be possible. 
The ${}^1S_0$ scattering lengths of the other potentials suggest also an
overall attractive interaction in this partial wave though only a very
moderate one. 

The results for the ${}^3S_1$ state of the $\Xi^0\Lambda$ channel are fairly
similar for all considered interactions. Moreover, with regard to the chiral EFT 
interaction there is very little cut-off dependence. 
The $S$-waves in the $\Xi\Sigma$ $I=3/2$ channel belong to the same 
($10^*$ and $27$, respectively) irreducible representations where in the 
$NN$ case real (${}^3S_1-{}^3D_1$) or virtual (${}^1S_0$) bound states exist,
cf. Table~\ref{tab:2.1}. Therefore, one expects that such states can also
occur for $\Xi\Sigma$. 
Indeed, bound states are present for both partial waves in
the Nijmegen model, cf. the discussion in Sect.~III.B in 
Ref.~\cite{Stoks:1999bz}. Their presence is reflected in the positive and
fairly large singlet and triplet scattering lengths for $\Xi^0\Sigma^+$, cf.
Table~\ref{tab:r2}. The chiral EFT interaction has positive scattering 
lengths of comparable magnitude for ${}^1S_0$, for all cut-off values, and
therefore bound states, too.  These binding energies lie in the range of 
$-2.23\,$MeV ($\Lambda = 550\,$MeV) to $-6.15\,$MeV (700 MeV). 
In the ${}^3S_1-{}^3D_1$ partial wave the
attraction is obviously not strong enough to form a bound state. The same
is the case (but for both $S$ waves) for the quark model fss2 of Fujiwara et al. 
\cite{Fujiwara:2006yh}.

The ${}^1S_0$ state of the $\Xi\Xi$ channel belongs also to the $27$plet
irreducible representation and also here the Nijmegen as well as the 
chiral EFT interactions produce bound states (see also \cite{Miller}). 
In our case the binding energies
lie in the range of $-2.56\,$MeV ($\Lambda = 550\,$MeV) to $-7.28\,$MeV 
(700 MeV). The predictions of both approaches for the ${}^3S_1$ scattering 
length are comparable. 
The quark model of Fujiwara et al. exhibits a different 
behavior for the $\Xi\Xi$ channel, see the last column in 
Table~\ref{tab:r2}. The small and negative ${}^1S_0$ 
scattering length signals an interaction that is only moderately attractive.
The large and positive scattering length in the ${}^3S_1-{}^3D_1$ 
partial wave, produced by that potential model, is usually a sign 
for the presence of a bound state, though according to the authors this 
is not the case for this specific interaction. 
Nevertheless, the $\Xi^0\Xi^-$ cross section predicted 
by the quark model is significantly larger than the results of our
chiral EFT interaction as well as those of the Nijmegen meson-exchange
potential, see, Table~\ref{tab:r1}.


\section{Summary}
\label{chap:5}

In this letter we have presented leading-order results for the $\Xi Y$ 
($Y = \Lambda, \Sigma$) and 
$\Xi\Xi$ interactions obtained within a chiral effective field theory approach based on 
Weinberg's power counting, derived analogous to the $YN$ potential presented in 
\cite{Polinder:2006zh}, by relating the strangeness $S=-3$ and $-4$ baryon-baryon 
interactions via ${\rm SU(3)}$ flavor symmetry to the one in the $YN$ system.

The LO chiral potential consists of two pieces: firstly, the longer-ranged 
one-pseudo{\-}scalar-meson exchanges and secondly, shorter-ranged four-baryon contact 
terms without derivatives.  
All the occuring five contact terms were already fixed in our study of the $YN$ interaction so
that genuine predictions can be made for the $S=-3$ and $-4$ baryon-baryon interactions
based on chiral EFT and the assumed ${\rm SU(3)_f}$ symmetry.
The reaction amplitude is obtained by solving a regularized coupled-channel 
Lippmann-Schwinger equation for the LO chiral potential. 
We used an exponential regulator function to regularize the potential and applied cut--offs 
in the range between $550$ and $700\,$MeV. 

It will be interesting to see whether the new facilities J-PARC (Tokai, Japan) and 
FAIR (Darmstadt, Germany) allow access to empirical information about the 
interaction in the $S=-3$ and $-4$ sector. Such information could come from
formation experiments of corresponding hypernuclei or from proton-proton and 
antiproton-proton collisions at such high energies that pairs of baryons with 
strangeness $S=-3$ or $S=-4$ can be produced. There is also a possibility to
find signals for strange di-baryon states in heavy-ion collisions 
\cite{Schaffner1999} which likewise would provide information on the 
corresponding baryon-baryon interaction. The chiral EFT developed here 
could then be used to analyze these upcoming data in a model-independent way.
It would be also interesting to see how these interactions can be tested in
multi-strange three-baryon systems recently explored in 
lattice QCD~\cite{Beane:2009gs}.

\vfill\eject
\ack

We would like to thank Andreas~Nogga for comments and a careful reading of our
manuscript. 
We acknowledge the support of the European Community-Research Infrastructure
Integrating Activity ``Study of Strongly Interacting Matter''
(acronym HadronPhysics2, Grant Agreement n. 227431)
under the Seventh Framework Programme of the EU.
Work supported in part by DFG (SFB/TR 16, ``Subnuclear Structure of Matter''),
and by the Helmholtz Association through funds provided to the virtual 
institute ``Spin and strong QCD'' (VH-VI-231) and by BMBF 
``Strong interaction studies for FAIR'' (grant 06BN9006).


\end{document}